\documentclass[reprint,aps,prb]{revtex4-1}
\pdfoutput=1
\usepackage{color,graphicx,calc,url}
\usepackage{dcolumn}
\usepackage{bm}
\usepackage{amssymb}
\usepackage{amsmath}
\usepackage{wasysym}
\usepackage{mathrsfs} 
\usepackage{array}
\usepackage{mathtools}
\usepackage{commath}
\usepackage{float}
\usepackage{textcomp}

\begin{document}
\title{Quadratic magnetooptic spectroscopy setup based on photoelastic light modulation}
\author{Robin Silber$^{1,2}$, Michaela Tom\'{i}\v{c}kov\'{a}$^{1}$, Jari Rodewald$^{3}$, Joachim Wollschl\"{a}ger$^{3}$, Martin Veis$^{4}$, Timo~Kuschel$^{2}$, Jaroslav Hamrle$^{1,4}$\email{Electronic mail: silber.rob@gmail.com}}
\affiliation{$^1$ Nanotechnology Centre, VSB-Technical University of Ostrava, 17. listopadu 15, 70833 Ostrava, Czech Republic\\
$^2$ Center for Spinelectronic Materials and Devices, Department of Physics, Bielefeld University, Universit\"atsstra\ss e 25, 33615 Bielefeld, Germany\\
$^3$ Department of Physics and Center of Physics and Chemistry of New Materials, Osnabr\"{u}ck University, 49076 Osnabr\"{u}ck, Germany\\
$^4$ Faculty of Mathematics and Physics, Charles University, Ke Karlovu 5, 12116 Prague, Czech Republic}

\date{\today}


\begin{abstract}

 In most of the cases the magnetooptic Kerr effect (MOKE) techniques rely solely on the effects linear in magnetization ($\bm{M}$). Nevertheless, a higher-order term being proportional to $\bm{M}$$^2$ and called quadratic MOKE (QMOKE) can additionally contribute to experimental data. Handling and understanding the underlying origin of QMOKE could be the key to utilize this effect for investigation of antiferromagnetic materials in the future due to their vanishing first order MOKE contribution. Also, better understanding of QMOKE and hence better understanding of magnetooptic (MO) effects in general is very valuable, as the MO effect is very much employed in research of ferro- and ferrimagnetic materials. Therefore, we present our QMOKE and longitudinal MOKE spectroscopy setup with a spectral range of 0.8--5.5\,eV. The setup is based on light modulation through a photoelastic modulator and detection of second-harmonic intensity by a lock-in amplifier. To measure the Kerr ellipticity an achromatic compensator is used within the setup, whereas without it Kerr rotation is measured. The separation of QMOKE spectra directly from the measured data is based on measurements with multiple magnetization directions. So far the QMOKE separation algorithm is developed and tested for but not limited to cubic (001) oriented samples. The QMOKE spectra yielded by our setup arise from two quadratic MO parameters $G_s$ and $2G_{44}$, being elements of quadratic MO tensor $\bm{G}$, which describe perturbation of the permittivity tensor in the second order in $\bm{M}$.
  
\end{abstract}

\maketitle

\section{Introduction}

The magnetooptic Kerr effect (MOKE)\cite{J.Kerr1877} technique is a tool that was and is vastly used for ferro- and ferrimagnetic material research. Setups employing MOKE at a single wavelength are typically used to provide information about magnetic properties of the sample such as magnetic anisotropy, magnetic remanence and coercivity, saturation field, magnetization ($\bm{M}$) reversal process or detection of exchange bias. On the other hand, setups providing MOKE response over a continuous spectrum (usually extended visible spectral range) yield the information about the electronic structure of the sample (here the results are usually accompanied by ab-initio calculations). \cite{Ferguson69, Krinchik68, Oppeneer92, Visnovsky1995, Uba1996, Visnovsky1999, Hamrle2001, Hamrle2002, Grondilova2002, Visnovsky2005, Veis2014, Sepulveda2003} Note that in both cases, one is usually relying on MOKE linear in $\bm{M}$ (LinMOKE), while its contribution quadratic in $\bm{M}$, quadratic MOKE (QMOKE), is considered to be more of a parasitic effect. Nevertheless, the QMOKE effects are often accompanying LinMOKE measurements, and hence its clear understanding is important. Further, the fact that QMOKE is even in $\bm{M}$ make it applicable for investigating antiferromagnetic materials which do not have a LinMOKE response.

At the beginning of the 90's unexpected symmetric contributions to the hysteresis loops of Ni-Fe bilayers were reported \cite{Zhong1990,Zhong1992} and later on explained as QMOKE contributions to the overall MOKE signal.\cite{Osgood1995, Osgood1997, Postava1997} Several methods have been proposed for the separation of QMOKE contributions from the LinMOKE signal including the ROTMOKE method,\cite{Mattheis1999} the 8-directional method,\cite{Postava2002} the sample rotation by 180$^\circ$,\cite{Mewes2004} and the rotation field method.\cite{Liang2015} Here we present QMOKE spectroscopy setup, which is capable to measure two types of QMOKE spectra in the spectral range of a 0.8--5.5\,eV, QMOKE$\sim G_s$ spectra and QMOKE$\sim 2G_{44}$, where $G_s$ and $2G_{44}$ are quadratic magnetooptic (MO) parameters that fully describe the perturbation to the permittivity tensor in the second order in $\bm{M}$ of materials with cubic crystallographic structure. \cite{Hamrle2007, Buchmeier2009, Kuschel2011, Hamrlova2013} LinMOKE spectra, namely longitudinal MOKE (LMOKE) spectra, can be measured as well. The measurement process is based on light modulation using photoelastic modulator and consequent detection by a lock-in amplifier. For the separation of LMOKE and of each QMOKE contribution, we employed a technique similar to the 8-directional method, but using a combination of just 4 directions and a sample rotation by 45$^\circ$ as described in this article.


\section{Theory of MOKE}
\label{theory_moke}

\subsection{Description of polarized light and MOKE}
MOKE manifests through the change of the polarization state of light reflected from a magnetized sample. Generally elliptically (fully) polarized light can be described by the Jones formalism using Jones vector such as \cite{Jones1941}

\begin{equation}
\textbf{J}_{\theta\epsilon}=
\begin{bmatrix}
\cos\theta\cos\epsilon-i\sin\theta\sin\epsilon\\[0.3em]
\sin\theta\cos\epsilon+i\cos\theta\sin\epsilon\\[0.3em]
\end{bmatrix}\,,
\label{generalpolar}
\end{equation}

\noindent 
where $\theta$ is the rotation angle of major axis of the polarization ellipse in our coordination system and $\epsilon$ is the ellipticity, i.e. arctan of the ratio of the minor and major axis of the polarization ellipse. Through those two values $\theta$ and $\epsilon$, an arbitrary state of full polarization can be described. The complex polarization parameter $\Phi$ is then defined as the ratio of the second and first component of the Jones vector $\bm{J}_{\theta\epsilon}$

\begin{equation}
\Phi=\frac{\sin\theta\cos\epsilon+i\cos\theta\sin\epsilon}{\cos\theta\cos\epsilon-i\sin\theta\sin\epsilon}=
\frac{\tan\theta+i\tan\epsilon}{1-i\tan\theta\tan\epsilon}\,.
\end{equation}

\noindent
In case of small angle approximation we can write 

\begin{equation}
\Phi=\theta+i\epsilon\,.
\label{small_angle}
\end{equation}

Withn Jones formalism, the reflection of polarized light from a sample is described by the reflection matrix\cite{Jones1941, Hecht2002}

\begin{equation}
	\bm{R}=
	\begin{bmatrix}
		r_{ss} & r_{sp}\\
		r_{ps} & r_{pp}
	\end{bmatrix}\,,
	\label{ref_matrix}
\end{equation} 

\noindent
where the first lower index of the elements refers to the polarization state of reflected light, while the latter index to the polarization state of the incident light. For isotropic systems without magnetization $r_{sp}=r_{ps}=0$. If the sample gets magnetized, a certain part of the reflected $s$-polarized wave is converted into a $p$-polarized wave (or vice versa). This change of polarization is described by the complex Kerr amplitude $\Phi_{s/p}$ for $s$ and $p$ polarized incident light. $\Phi_{s/p}$ is actually analogous to the complex polarization parameter $\Phi$ from Eq.~(\ref{small_angle}). With respect to the Jones formalism and  Eq.~(\ref{ref_matrix})\cite{Jones1941, Visnovsky06}

\begin{equation}
\begin{split}	
\Phi_{s}&{}=\theta_{s}+i\epsilon_{s}=
-\frac{r_{ps}}{r_{ss}}\,,
\\[5mm]
\Phi_{p}&{}=\theta_{p}+i\epsilon_{p}=
\frac{r_{sp}}{r_{pp}}
\,.
\end{split}
\label{Kerr_basic}
\end{equation}

\noindent
Here $\theta_{s/p}$ is called Kerr rotation and $\epsilon_{s/p}$ is called Kerr ellipticity and in common called Kerr angles. Note that these Kerr angles are usually smaller than 1 degree, and therefore use of  small angle approximation is appropriate.

\subsection{Theory of linear and quadratic MOKE}
The reflection coefficients from Eq.~(\ref{Kerr_basic}) are strictly bound with the permittivity tensor $\bm{\varepsilon}$ of the crystal. The permittivity tensor elements of ferromagnetic (FM) material, 
\begin{equation}
	\varepsilon_{ij}=\varepsilon_{ij}^{(0)}+K_{ijk}M_{k}+G_{ijkl}M_kM_l\,,
	\label{permitivity_FM}
\end{equation} 

\noindent
are described up to second order in magnetization $\bm{M}$ by the permittivity in the 0th order in $\bm{M}$ ($\bm{\varepsilon}^{(0)}$) and by the linear and quadratic MO tensor $\bm{K}$ and $\bm{G}$, respectively.\cite{Visnovsky1986} $M_k$ and $M_l$ are components of normalized $\bm{M}$. In case of the cubic crystal structure (classes $\bar{4}3m$, $432$, $m3m$), the linear MO tensor is described by one free parameter $K$ and the quadratic MO tensor is described by two free parameters $2G_{44}$ and $G_s=(G_{11}-G_{12})$.\cite{Visnovsky1986, Visnovsky06, Hamrlova2013} The permittivity tensor in the 0th order in $\bm{M}$ is described by the scalar $\varepsilon_{(d)}$, $\bm{\varepsilon}^{(0)}=\varepsilon_{(d)}\bm{I}$, where $\bm{I}$ is identity matrix. 

The analytical approximation of MOKE response for FM layers is \cite{Hamrle2007}

\begin{equation}
\begin{split}
\Phi_s &{}=-\frac{r_{ps}}{r_{ss}}= A_{s}\left(\varepsilon_{yx}-\frac{\varepsilon_{yz}\varepsilon_{zx}}{\varepsilon_{(d)}}\right)+B_s\varepsilon_{zx}\,,\\[5mm]
\Phi_p &{}=\frac{r_{sp}}{r_{pp}}=-A_{p}\left(\varepsilon_{xy}-\frac{\varepsilon_{zy}\varepsilon_{xz}}{\varepsilon_{(d)}}\right)+B_p\varepsilon_{xz}\,,
\end{split}
\label{Kerr_analyt}
\end{equation}

\noindent
where $A_{s/p}$ and $B_{s/p}$ are the optical weighting factors. If we will limit ourselves to in-plane normalized $\bm{M}$ and cubic crystal structure with [001] direction normal to the surface, then the dependence of the MOKE amplitude on the MO parameters $K$, $G_s$, $2G_{44}$, on the sample orientation $\alpha$ and on the in-plane $\bm{M}$ direction $\mu$ can be derived from Eqs.~(\ref{permitivity_FM}) and (\ref{Kerr_analyt}),\cite{Hamrle2007, Kuschel2011, Silber2014} resulting in

\begin{equation}
\begin{split}	
\Phi_{s/p} =&{}\pm A_{s/p}\left\{\frac{2G_{44}}{4}\left[\left(1+\cos{4\alpha}\right)\sin{2\mu}+\sin{4\alpha}\cos{2\mu}\right]\right.\,  \\[4mm]
&{}\qquad\quad+\frac{G_{s}}{4}\left.\left[\left(1-\cos{4\alpha}\right)\sin{2\mu}-\sin{4\alpha}\cos{2\mu}\right]\frac{}{}\right\} \\[4mm]
&{}\mp A_{s/p}\,\,\frac{K^2}{2\varepsilon_{(d)}}\sin{2\mu} \\[4mm]
&{}\pm B_{s/p}\,\,K\sin{\mu}\,.
\end{split}
\label{KerrDepend}
\end{equation}


\noindent
The sign $\pm$ is given by the polarization ($s/p$) of the incident light beam. This dependence provides us with measurement sequences, which isolate the individual MOKE contributions that have its origin in the individual MO parameters.\cite{Postava2002} These measurement sequences with exact definitions of the angles $\alpha$ and $\mu$ will be discussed in Sec.~\ref{subsec_conventions} and \ref{subsec_measurementprinc}.


\section{Setup description}
\subsection{Conventions and definitions}
\label{subsec_conventions}

At first, we have to introduce conventions for positive and negative rotations of crystallographic structure, of $\bm{M}$, of light polarization and of optical elements. We further have to define all coordinate systems used within the setup in which those rotations take place. To describe reflection on the sample, three cartesian systems are needed, one for the incident light beam, one for the reflected light beam and one for the sample.

(i) The electric field vector of an electromagnetic wave is described by a negative time convention as $\bm{E}{(\bm{r},t)}=\bm{E}(\bm{r}) e^{-i\omega t}$, providing permittivity in form of $\varepsilon=\varepsilon_R +i\varepsilon_I$, where  $\varepsilon_{R}$, $\varepsilon_{I}$ being real, imaginary part of complex permittivity, respectively, where $ \varepsilon_I >0$.

\begin{figure}
\begin{center}
\includegraphics[width=0.45\textwidth]{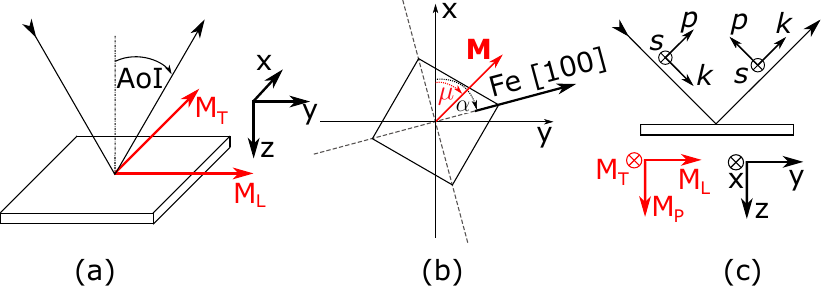}
\end{center}
\caption{(a) Right-handed coordinate system $\hat{x}$, $\hat{y}$, $\hat{z}$ is established  with respect to the plane of incidence and surface of the sample. Components of the in-plane normalized $\bm{M}$ $M_{T}$, $M_{L}$ are defined along axis $\hat{x}$, $\hat{y}$ of the coordinate system, respectively. The AoI stands for angle of incidence. (b) Definition of positive in-plane rotation of the sample and $\bm{M}$ within the $\hat{x}$, $\hat{y}$, $\hat{z}$ coordinate system, described by angle $\alpha$ and $\mu$, respectively. (c) Definition of the right-handed cartesian system $\hat{s}$, $\hat{p}$, $\hat{k}$ of incident and reflected beam. All directions and angles shown in the figure are of positive values}
\label{xyz_sys}
\end{figure}

(ii) The cartesian system describing the sample is the right-handed $\hat{x}$, $\hat{y}$, $\hat{z}$ system, where $\hat{z}$-axis is normal to the surface of the sample and points into the sample. The $\hat{y}$-axis is  parallel with the plane of light incidence and with the sample surface, while its positive direction is defined by the direction of $k_y$, being the $\hat{y}$-component of the wave vector of incident light as shown in Fig.~\ref{xyz_sys}. In this system, rotations of the crystallographic structure and $\bm{M}$ take place.

(iii) We use a right-handed cartesian system $\hat{s}$, $\hat{p}$, $\hat{k}$ for description of the incident and reflected light beam. The direction of vector $\hat{k}$ defines the direction of the propagation of light. Vector $\hat{p}$ lies in the plane of incidence, i.e.\ a~plane defined by the incident and reflected beam. Vector  $\hat{s}$ is perpendicular to this plane and corresponds to $\hat{x}$.  This convention is the same for both incident and reflected beam (Fig.~\ref{xyz_sys}).
	
(iv) The Kerr rotation $\theta$  is positive if azimuth $\theta$ of the polarization ellipse  rotates clockwise, when looking into the incoming light beam. The Kerr ellipticity $\epsilon$ is positive if the electric field vector $\bm{E}$ rotates clockwise when looking into the incoming light beam.
	
(v) The angle of rotation of the sample, optical elements and the magnetization $\bm{M}$ is defined as positive, if the rotated vector pointing in $\hat{x}$ ($\hat{s}$) direction rotates towards $\hat{y}$ ($\hat{p}$) direction. Hence, the sample orientation $\alpha$=0 corresponds to the [100] direction in cubic crystals being parallel to the $\hat{x}$-axis and, when looking at the top surface of the sample, the positive rotation of the sample is clockwise. The same applies to the in-plane $\bm{M}$ direction described by angle $\mu$. Further, when looking into the incoming beam, the positive rotation of the optical elements is counter-clockwise, being in contrast to the Kerr angles, originating from the historical convention of the ellipsometric and MO angles.

\subsection{Elements of the setup}
The source of light is provided by a Xenon short arc lamp (extended to UV region, 300 W) followed by a grid monochromator (Oriel Cornerstone 260 1/4 m) in Czerny-Turner optical configuration. A Rochon prism beam-splitting polarizer is then used to yield $s$-polarized or $p$-polarized incident waves. The sample is exposed to an in-plane magnetic field, provided by a magnetic circuit with permanent magnets (300mT), which can be rotated by a rotational stage by an arbitrary angle $\mu$. The rotational stage of the sample holder provides a precise rotation of the sample by an arbitrary angle $\alpha$. After reflection from the magnetized sample, the light travels through an optional optical element - achromatic compensator - providing a phase shift of $\delta$=90$^\circ$. When the compensator is present (absent) the setup measures Kerr ellipticity (rotation). The light further propagates through a photoelastic modulator (PEM) (Hinds Instruments PEM-100) bound with an analyzer (Rochon prism) at 45$^\circ$. Afterwards, one of the three detectors, being infrared diode (Newport 7032 8NS) or photomultipliers for visible (Hamamatsu H7712-13) or ultra-violet (Hamamatsu H9307) region of spectra, detects the reflected light, respectively. To manipulate the light beam, parabolic mirrors are used through the whole setup and hence our setup is completely chromatic-aberration free. The spectral range of the setup (determined by the spectral characteristics of lamp, monochromator and detectors) is 0.8--5.5\,eV.

The signal from the detector is then processed by a lock-in amplifier (Stanford Research System SR830), with a reference frequency being the frequency of the PEM. The setup is controlled via an in-house written code in python 2.7 language, using the pyVisa interface to communicate with the hardware of the setup. 

\begin{figure}
\begin{center}
\includegraphics[width=0.45\textwidth]{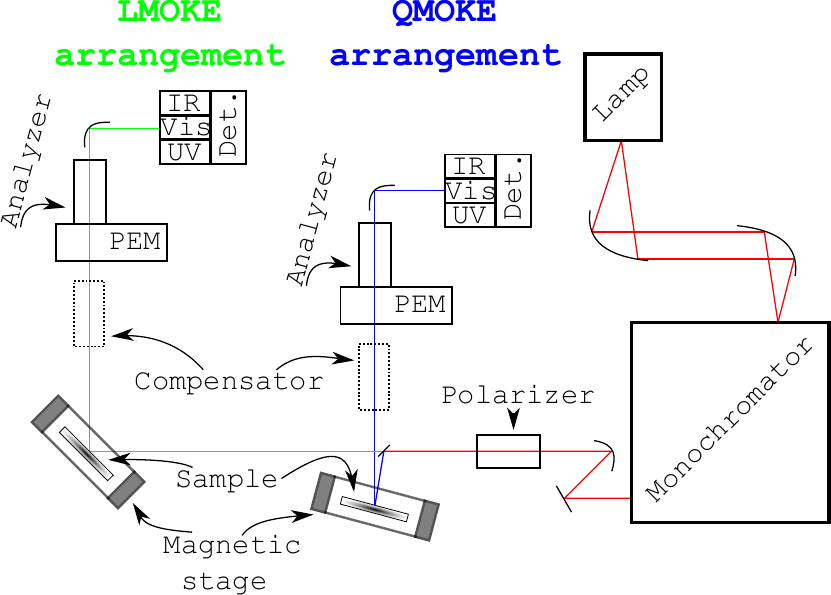}
\end{center}
\caption{Sketch of the optical elements of the setup on the optical table. The optical elements are: lamp--monochromator--polarizer--sample with magnetic stage--(compensator)--PEM and analyzer--detectors. The optical path denoted by the red beam is stable for both, LMOKE and QMOKE configuration. To measure LMOKE, the green path is used while to measure QMOKE the blue  path have to be arranged.}
\label{setup_sketch}
\end{figure}

\subsection{MOKE measurement principle}
\label{subsec_measurementprinc}

The the polarization state of light propagation through the setup is described in the Jones formalism as

\begin{equation}
	\bm{J}_{\mathrm{det}}= \bm{AP}(\bm{C}^{(\frac{\pi}{2})})\bm{RJ}_{\mathrm{in}}\,,
\end{equation}

\noindent
which can be written in matrix form as

\begin{equation}
\begin{split}
\begin{bmatrix}
E_s\\
E_p\\
\end{bmatrix}
=
E_0
\begin{bmatrix}
1 & 1\\
1 & 1\\
\end{bmatrix}
\begin{bmatrix}
e^{i\frac{\varphi}{2}} &   0 \\
0  &  e^{-i\frac{\varphi}{2}} \\
\end{bmatrix}
\begin{bmatrix}
\cos{\beta} & \sin{\beta}\\
-\sin{\beta} & \cos{\beta}\\
\end{bmatrix}
\\
\left(
\begin{bmatrix}
e^{i\frac{\pi}{4}} &   0 \\
0  &  e^{-i\frac{\pi}{4}} \\
\end{bmatrix}
\right)
\begin{bmatrix}
r_{ss} & r_{sp}\\
r_{ps} & r_{pp}\\
\end{bmatrix}
\begin{bmatrix}
\cos\xi \\
\sin\xi \\
\end{bmatrix}\,.
\label{second_arrang}
\end{split}
\end{equation}

\noindent
Here, $\bm{J}_{\mathrm{det}}$ describes the electric field amplitudes at the detector. $\bm{A}$ is the Analyzer  at $45^\circ$ with PEM ($\bm{P}$). Phase of modulation $\varphi=\varphi_S+\varphi_A\sin{(\omega t)}$, where $\varphi_S$ is phase shift constant in time, $\varphi_A$ is the modulation amplitude and $\omega$ is frequency of PEM. This bound $\bm{AP}$ optical element can be rotated by an angle $\beta$. $\bm{C}^{(\frac{\pi}{4})}$ is the optional optical element - achromatic quarter-wave compensator. $\bm{R}$ is the reflection matrix of the sample (Eq.~(\ref{ref_matrix})) and $\bm{J}_{\mathrm{in}}$ is the Jones vector of the incident light given by the polarizer at angle $\xi$ and being $\left[1,0\right]$, $\left[0,1\right]$ for $s$, $p$ polarized incident light, respectively. $E_0$ is a constant prefactor and its absolute value is not important for our investigation.

At the detector, the intensity is measured. As analyzer is oriented at $45^\circ$, hence $E_s=E_p$ and $E=\sqrt{E_s^2+E_p^2}$, we can introduce the overall electric field intensity as $E=\sqrt{2}E_s=\sqrt{2}E_p$. Hence, we can write the intensity at the detector as $I=2I_0|E_s|^2$, with $I_0$ being the intensity prefactor.

The following steps and approximations are made when we analyze the intensity at the detector: (\textit{i}) We apply small angle approximations for all the Kerr angles and for the rotation angle $\beta$ of the $\bm{AP}$ optical element. (\textit{ii}) We neglect all the terms with square of the Kerr angle. (\textit{iii}) We expand $e^{i\varphi}$ into  Bessel functions.\cite{Arfken95} From PEM calibration we got $\varphi_S\approx 0$ and with use of small angle approximation we can write $\sin{\varphi_S}=0$ and $\cos{\varphi_S}=1$. Then, $e^{i\varphi}=J_0(\varphi_A)+i2J_1(\varphi_A)\sin{(\omega t)}+2J_2(\varphi_A)\cos{(2\omega t)}$. For simplicity, we show here only the second-harmonic intensity $I_{2\omega}$ measured by the lock-in amplifier, as it solely is enough to conduct the MOKE measurements. For more detailed calculations, see the literature.\cite{Silber2014} $I_{2\omega}$ at the detector for $s$-polarized incident beam with and without compensator and for $p$- polarized incident beam with and without compensator is

\begin{table}
\begin{center}
\begin{tabular}{||p{0.2\textwidth}|c|c||}\hline
\multicolumn{3}{||c||}{\bfseries Measurement techniques} \rule{0pt}{1.3em}\\ [0.3em]
\cline{1-3}
 Measured Kerr effect &
 $\Phi_s$& \rule{0pt}{1.7em}
$\Phi_p$ \rule{0pt}{1.7em}\\ [2mm]
\hline
Polarizer orientation&
$\xi=0$ \rule{0pt}{1.7em} &
$\xi=\frac{\pi}{2}$\rule{0pt}{1.7em}\\ [2mm]
\cline{1-3}
Rotation measurement &
$\theta_s=\frac{\Delta I_{2\omega}^{(s)}}{2\gamma^{(s)}}$ \rule{0pt}{1.7em}& 
$\theta_p=\frac{\Delta I_{2\omega}^{(p)}}{2\gamma^{(p)}}$ \rule{0pt}{1.7em}\\ [2mm]
 \hline
Ellipticity measurement &
$\epsilon_s=\frac{\Delta I_{2\omega}^{(s,c)}}{2\gamma^{(s,c)}}$ \rule{0pt}{1.7em} & 
$\epsilon_p=-\frac{\Delta I_{2\omega}^{(p,c)}}{2\gamma^{(p,c)}}$  \rule{0pt}{1.7em}\\ [2mm]
\hline
\end{tabular}
\end{center}
\caption{The measurement techniques for the setup with arrangements of optical elements: polarizer--sample--(compensator)--PEM and analyzer--detector. The calibration slope $\gamma^{(s)/(s,c)/(p)/(p,c)}$ is obtained from a calibration measurement provided by precise PEM+analyzer rotation.}
\label{measurement_sum_set2}
\end{table}


\begin{subequations}
\begin{alignat}{4}
\mathmakebox[\widthof{QQQ}][l]{I_{2\omega}^{(s,c)}}&=&\mathmakebox[\widthof{QQQQQQQQQQQ}][r]{-I_k|r_{ss}|^2\left(\epsilon_s+\beta_s\right)}&=&\mathmakebox[\widthof{QQQQQQQQQQ}][c]{\gamma^{(s,c)}\left(\epsilon_s+\beta\right)\,,}
\\[3mm]
\mathmakebox[\widthof{QQQ}][l]{I_{2\omega}^{(s)}}&=&\mathmakebox[\widthof{QQQQQQQQQQQ}][r]{-I_k|r_{ss}|^2\left(\theta_s+\beta_s\right)}&=&\mathmakebox[\widthof{QQQQQQQQQQ}][c]{\gamma^{(s)}\left(\theta_s+\beta\right)\,,}
\\[3mm]
\mathmakebox[\widthof{QQQ}][l]{I_{2\omega}^{(p,c)}}&=&\mathmakebox[\widthof{QQQQQQQQQQQ}][r]{-I_k|r_{pp}|^2\left(\epsilon_p-\beta_p\right)}&=&\mathmakebox[\widthof{QQQQQQQQQQ}][c]{\gamma^{(p,c)}\left(\epsilon_p-\beta\right)\,,}
\\[3mm]
\mathmakebox[\widthof{QQQ}][l]{I_{2\omega}^{(p)}}&=&\mathmakebox[\widthof{QQQQQQQQQQQ}][r]{I_k|r_{pp}|^2\left(\theta_p+\beta_p\right)}&=&\mathmakebox[\widthof{QQQQQQQQQQ}][c]{\gamma^{(p)}\left(\theta_p+\beta\right)\,,}
\end{alignat}
\end{subequations}

\noindent
where superscript $(c)$ denotes the presence of the compensator in the setup, $\omega$ is modulation frequency of the PEM and $I_k=4J_2(\varphi_A)k_{2\omega}I_0$, where $k_{2\omega}$ is electronic transmission coeficient for $2\omega$ frequency. In all four cases, the intensity at 2$\omega$ has a linear dependence on $\gamma$ that is the same for the Kerr angle and angle $\beta$. Hence, the Kerr angles can be easily measured by a change of $I_{2\omega}$ with $\bm{M}$ and by the knowledge of absolute value of linear dependence slope $\gamma$ (volt-degree conversion factor), which is obtained by the $\bm{AP}$ optical element rotation by small angle $\beta$. It's important to keep in mind, that this volt-degree conversion factor $\gamma$ is unique for each wavelength and sample orientation. The measurement method for both polarizations and for both Kerr angles are summarized in Tab.~\ref{measurement_sum_set2}.

Now, with use of Eq.~(\ref{KerrDepend}) we can develop the MOKE measurements processes that separate  linear and quadratic contributions directly from the measured data. The separation is based on MOKE measurement with different $\bm{M}$ direction and sample orientation. \cite{Postava2002}

\begin{widetext}
\begin{subequations}
\begin{alignat}{4}
&\mathmakebox[\widthof{QQQQQQQQQQQQ}][l]{\mathrm{LMOKE}\sim K:}&
\frac{I_{2\omega}^{\mu=90^{\circ}}-I_{2\omega}^{\mu=270^{\circ}}}{2\gamma^{(s/p,\,c/\,\,\,)}}
&\mathmakebox[\widthof{EEE}][c]{=}&
&\mathmakebox[\widthof{AAAAAAAAAAA}][l]{\pm\,B_{s/p}K ,}&
\begin{array}{r@{}l}
	&{}\alpha = \mathrm{arb.\,angle}\\ 
	&{}\mathrm{AoI}=45^{\circ}
\end{array}\,.
\label{LMOKE_sequence}
\\[3mm]
&\mathmakebox[\widthof{QQQQQQQQQQQQ}][l]{\mathrm{QMOKE}\sim G_s:}&
\frac{I_{2\omega}^{\mu=45^{\circ}}+I_{2\omega}^{\mu=225^{\circ}}-I_{2\omega}^{\mu=135^{\circ}}-I_{2\omega}^{\mu=315^{\circ}}}{2\gamma^{(s/p,\,c/\,\,\,)}}
&\mathmakebox[\widthof{EEE}][c]{=}&
&\mathmakebox[\widthof{AAAAAAAAAAA}][l]{\pm\,A_{s/p}\left(G_{s}-\frac{K^2}{\varepsilon_{(d)}}\right) ,}&
\begin{array}{r@{}l}
	&{}\alpha = 45^{\circ}\\
	&{}\mathrm{AoI}=5^{\circ}
\end{array}\,.
\label{QMOKE_sequence_s}
\\[3mm]
&\mathmakebox[\widthof{QQQQQQQQQQQQ}][l]{\mathrm{QMOKE}\sim 2G_{44}:}&
\frac{I_{2\omega}^{\mu=45^{\circ}}+I_{2\omega}^{\mu=225^{\circ}}-I_{2\omega}^{\mu=135^{\circ}}-I_{2\omega}^{\mu=315^{\circ}}}{2\gamma^{(s/p,\,c/\,\,\,)}}
&\mathmakebox[\widthof{EEE}][c]{=}&
&\mathmakebox[\widthof{AAAAAAAAAAA}][l]{\pm\,A_{s/p}\left(2G_{44}-\frac{K^2}{\varepsilon_{(d)}}\right) ,}& 
\begin{array}{r@{}l}
	&{}\alpha = 0^{\circ}\\
	&{}\mathrm{AoI}=5^{\circ}
\end{array}\,.
\label{QMOKE_sequence_44}
\end{alignat}
\end{subequations}
\end{widetext}

\noindent
Although another angle of incidence (AoI) could be used within those measurement sequences as well, we choose AoI presented in Eqs.~(\ref{LMOKE_sequence})--(\ref{QMOKE_sequence_44}) because of technical reasons. On the other hand, $\bm{M}$ direction $\mu$ and sample orientation $\alpha$ are crucial to this MOKE separation measurement process.


\section{Measurements}

The measurement processes described in Eqs.~(\ref{LMOKE_sequence})--(\ref{QMOKE_sequence_44}) were applied on Fe$_{3}$O$_{4}$ (magnetite) thin film samples grown on MgO(001) substrates by molecular beam epitaxy.\cite{Scheme2015}

\begin{figure}
\begin{center}
\includegraphics{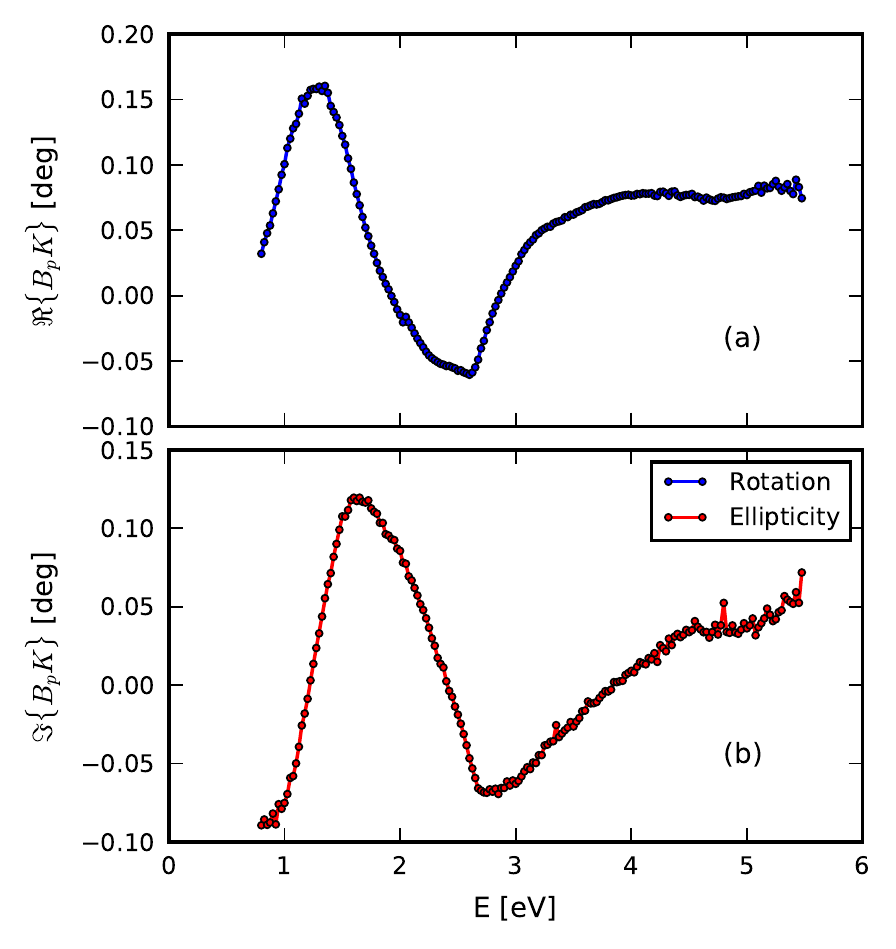}
\end{center}
\caption{LMOKE spectra measured according to Eq.~(\ref{LMOKE_sequence}). (a) LMOKE rotation, (b) LMOKE ellipticity. }
\label{LMOKE_spec}
\end{figure}

\begin{figure}
\begin{center}
\includegraphics{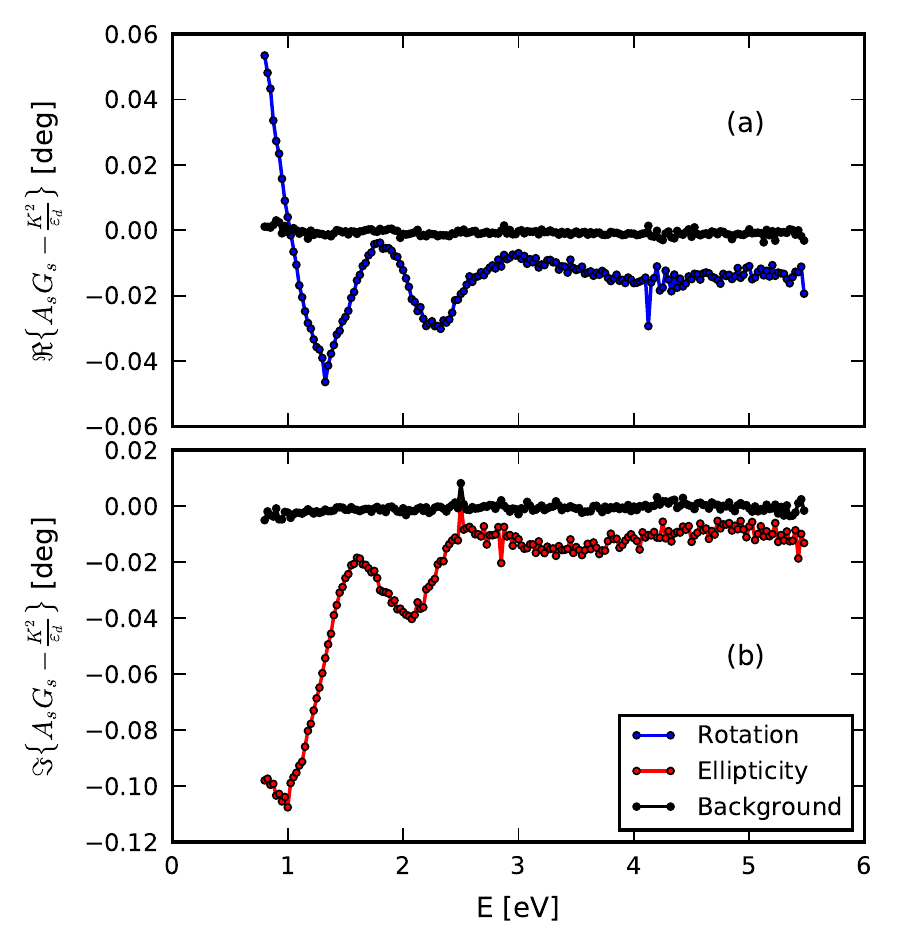}
\end{center}
\caption{QMOKE spectra measured according to Eq.~(\ref{QMOKE_sequence_s}). (a) QMOKE$\sim G_{s}$ rotation (b) QMOKE$\sim G_s$ ellipticity. Background spectra should be noise-around zero signal.}
\label{QMOKE_Gs_spec}
\end{figure}

\begin{figure}
\begin{center}
\includegraphics{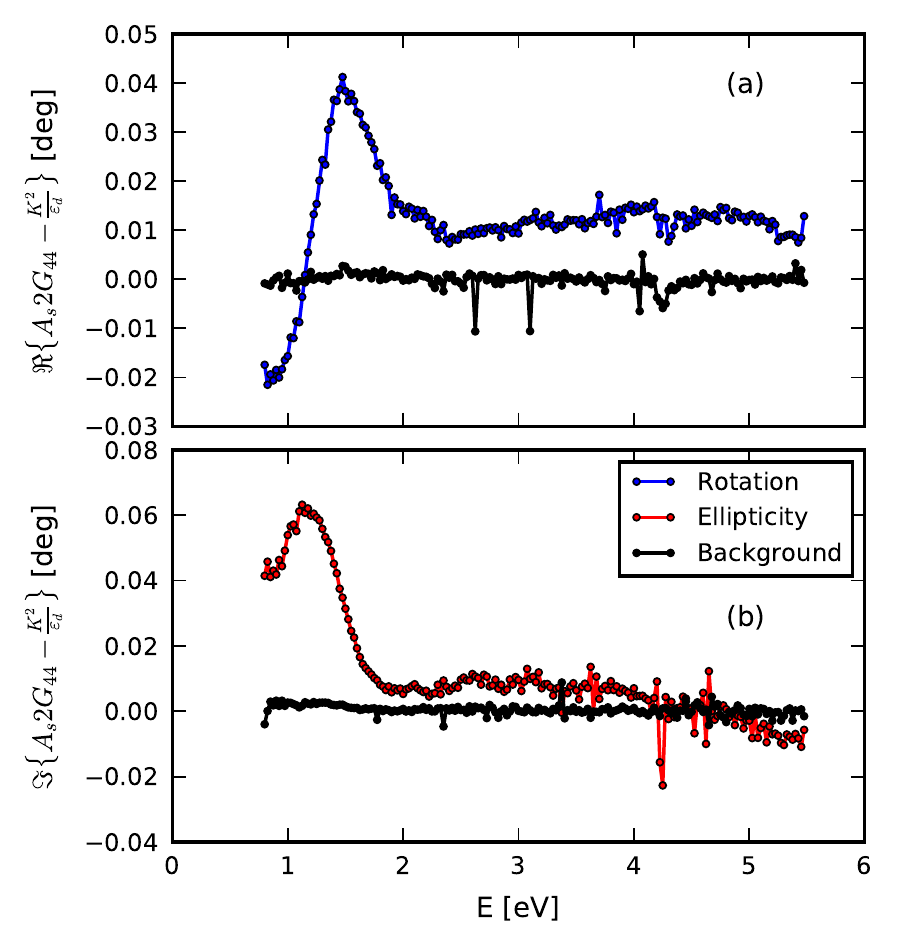}
\end{center}
\caption{QMOKE spectra measured according to Eq.~(\ref{QMOKE_sequence_44}). (a) QMOKE$\sim 2G_{44}$ rotation (b) QMOKE$\sim 2G_{44}$ ellipticity. Background spectra should be noise-around zero signal.}
\label{QMOKE_2G44_spec}
\end{figure}

The LMOKE spectra measured according to Eq.~(\ref{LMOKE_sequence}) are presented in Fig.~\ref{LMOKE_spec} with two clearly visible peaks at  1.3\,eV and 2.5\,eV in the LMOKE rotation. The QMOKE spectra measured according to Eq.~(\ref{QMOKE_sequence_s}) are presented in Fig.~\ref{QMOKE_Gs_spec} with multiple peaks in range 0.8\,eV--3\,eV and the QMOKE spectra measured according to Eq.~(\ref{QMOKE_sequence_44}) are presented in Fig.~\ref{QMOKE_2G44_spec} with single well pronounced peak at 1.3\,eV. The peaks in the Kerr ellipticity spectra and its shape in general are interconnected with the Kerr rotation spectra by Kramers-Kronig relations. The peaks in the spectra could be assign to certain transitions in the material known from literature,\cite{Zviagin2016} or compared to ab-initio calculation for more detailed treatment. The detailed description and interpretation of the presented LMOKE and QMOKE spectra is out of scope of this paper and will be presented elsewhere.

 In the latter two figures, we also show a so-called background signal, being the signal measured through the same manner as the QMOKE signal (Eqs.~(\ref{QMOKE_sequence_s}) and (\ref{QMOKE_sequence_44})) but the in-plane $\bm{M}$ directions are $\mu=0^{\circ},90^{\circ},180^{\circ},270^{\circ}$. Such a measurement sequence should provide zero MO response, as one can easily read out from Eq.~(\ref{KerrDepend}). Hence, as the only difference between QMOKE spectra and background spectra is the different $\bm{M}$ directions in the measurement sequence, it is the perfect way to test the setup for artefacts. The results shown in Figs.~\ref{QMOKE_Gs_spec} and \ref{QMOKE_2G44_spec} approve the correctness of the measurement sequences for the QMOKE spectra based on the theory (Eq.~(\ref{KerrDepend})). On the other hand the background spectra are not strictly noise-around-zero signal. This non-zero signal in the background spectra can be explained by (\textit{i}) non-ideal cubic sample (non-ideal epitaxial growth on MgO substrate with some secondary preferable grow direction) (\textit{ii}) slight misalignment of the sample in the setup, i.e. misalignment of [100] direction of the sample with respect to $\alpha=0^\circ$ and (\textit{iii}) artefacts of the setup.
 
  Although the measurement process used within this setup eliminates most of the possible artefacts (detection of change of $I_{2\omega}$ intensity with $\bm{M}$ - absolute value of $I_{2\omega}$ intensity is not important), the change of $\bm{M}$ direction is provided by a rotational stage with permanent magnets in a magnetic circuit. Therefore, the shade of this magnetic stage is not identical for different magnetization directions (compared to an electromagnet, which does have identical shade for all magnetization directions) and hence, this could result in different level of scattered light entering the detector. This effect can be more pronounced at certain wavelengths, because amount of scattered light from various surfaces, sample and optical elements will be wavelength dependent. This is also most probably the origin of a slight peak in Fig.~\ref{QMOKE_2G44_spec}(a) at 4--4.5eV that has same magnitude for background and QMOKE spectra. The elimination of this artefact can be done by proper centring of the sample (i.e. sample holder) in the magnetic stage and by eliminating all the reflections of extraordinary beams from polarizer and analyzer.


\section{Conclusion}

We introduced our in-house built LMOKE and QMOKE spectroscopy setup that operates in a spectral range of 0.8--5.5\,eV. The setup is based on the detection of change of second harmonic intensity (light modulated by PEM) with change of $\bm{M}$ of the sample. The polarization measurement algorithm of QMOKE spectra is applicable on the samples with cubic structure and (001) oriented surface. Measurement sequences for different orientations and different crystallographic structures could be developed from shape of permittivity tensor that is described up to the second order in $\bm{M}$. The QMOKE and LMOKE spectra were measured on magnetite (Fe$_{3}$O$_{4}$) thin film samples grown on MgO(001) substrates by molecular beam epitaxy. The precision of the setup is good enough to observe well recognisable features in the QMOKE spectra. The QMOKE spectra, when compared to the LMOKE spectra, provide clearly an additional information, hence the obtained data are suitable for further analysis. Through the measurement of the so-called background signal, the signal that should be noise-around-zero from theory, we check the correctness of the measurement sequences and also test the setup for possible artefacts.

\begin{acknowledgments}
This work was supported by the European Regional Development Fund in the IT4Innovations national supercomputing center - path to exascale project, project number CZ. $02.1.01/0.0/0.0/16\_013/0001791$ within the Operational Programme Research, Development and Education.


\end{acknowledgments}

\newpage
\bibliography{QMOKE_setup}

\end{document}